\documentclass[%
reprint,
superscriptaddress,
amsmath,amssymb,
aps,
prr,
floatfix,
]{revtex4-1}

\usepackage{graphicx}
\usepackage{dcolumn}
\usepackage{bm}
\usepackage{braket}
\usepackage[german,english]{babel}
\usepackage[colorlinks=true,linkcolor=blue,citecolor=blue,breaklinks]{hyperref}
\usepackage[pdftex]{color}
\usepackage{ulem}
\usepackage{comment}
\usepackage{natbib}
\usepackage{xcolor}
\usepackage{amsmath}
\DeclareMathOperator{\Tr}{Tr}
\usepackage{tikzsymbols}
\begin{document}

	\title{Visualizing Quasiparticles from Quantum Entanglement for general 1D phases}
		\author{Elisabeth Wybo}
\affiliation{Department of Physics, Technical University of Munich, 85748 Garching, Germany}
	\affiliation{Munich Center for Quantum Science and Technology (MQCST), D-80799 Munich, Germany}
\author{Frank Pollmann}
\affiliation{Department of Physics, Technical University of Munich, 85748 Garching, Germany}
	\affiliation{Munich Center for Quantum Science and Technology (MQCST), D-80799 Munich, Germany}
	\author{S. L. Sondhi}
	\affiliation{Department of Physics, Princeton University, NJ 08544, USA}
	\author{Yizhi You}
	\affiliation{Department of Physics, Princeton University, NJ 08544, USA}	

	\date{\today}
	
	\begin{abstract}
	In this work, we present a quantum information framework for the entanglement behavior of the low energy quasiparticle (QP) excitations in various quantum phases in one-dimensional (1D) systems. We first establish an exact correspondence between the correlation matrix and the QP entanglement Hamiltonian for free fermions  and find an extended in-gap state in the QP entanglement Hamiltonian as a consequence of the position uncertainty of the QP. A more general understanding of such an in-gap state can be extended to a Kramers theorem for the QP entanglement Hamiltonian, which also applies to strongly interacting systems.
	Further, we present a set of ubiquitous entanglement spectrum features, dubbed entanglement fragmentation, conditional mutual information, and measurement induced non-local entanglement for QPs in 1D symmetry protected topological phases. Our result thus provides a new framework to identify different phases of matter in terms of their QP entanglement. 
\end{abstract}

\maketitle

\section{Introduction}

Considerable progress has been made over the past decades in elucidating the nature of the entanglement behavior in many-body systems from a quantum information perspective~\cite{RevModPhys.80.517,calabrese2008entanglement,calabrese2009entanglement,Peschel_2009,pollmann2012symmetry,Fannes-1992,Verstraete9,Schollwoeck11,PhysRevB.84.195103,PhysRevLett.105.115501,PhysRevLett.104.130502,chandran2014universal,PhysRevLett.110.236801,zhang2012quasiparticle}. A prominent example is the topological entanglement entropy (EE) whose scaling exhibits a correction to the area law as a manifestation of topological order~\cite{KitaevPreskill,Levin-2006}. Following this identification, it was proposed that symmetry protected topologically phases (SPT) can be characterized by the presence of “fractionalized” modes in the entanglement spectrum (ES), i.e., the spectrum of the reduced density matrix (RDM) $\rho_A$, which transform differently under the symmetry group from the constituent microscopic degrees of freedom of the system~\cite{chen2010local,chen2012symmetry,Chen2011-kz,Chen2011-et,Pollmann2010,Pollmann2012-lv,pollmann2012symmetry,Turner2011-zi,Li-2008,Peschel_2009,pollmann2012symmetry,PhysRevB.84.195103,PhysRevLett.105.115501,PhysRevLett.104.130502,chandran2014universal,PhysRevLett.110.236801}. The relation between the topological structure and the ES has been widely explored since then~\cite{Li-2008,Peschel_2009,pollmann2012symmetry,PhysRevB.84.195103,PhysRevLett.105.115501,PhysRevLett.104.130502,chandran2014universal,PhysRevLett.110.236801}. Remarkably, most salient topological properties, including quasiparticle statistics, edge excitations, central charge and topological Berry phase can be readily reached by scrutinizing the entanglement spectrum~\cite{Li-2008,zhang2012quasiparticle,tu2013momentum,matsuura2016charged,marvian2017symmetry}. Such correspondences provide a powerful tool to explore quantum phases or critical points from a quantum information perspective, and have been eminently successful in describing a wide variety of exotic states or phase transition phenomena~\cite{calabrese2008entanglement,ding2009entanglement,zhang2012quasiparticle}.

While the entanglement structure for the quantum many-body ground state has been widely explored, much less is known for the quasiparticle (QP) excitation at low energy~\cite{metlitski2011entanglement,castro2018entanglement,eisler2019front,castro2019entanglement,jafarizadeh2019bipartite,castro2018entanglementa}. In particular, the universal entanglement features of low energy quasiparticle states in diverse phases remain unclear and leave important open questions. It is widely accepted that the ground state wave-function encodes essential topological features of the phase; does the QP excitation still inherit similar features? Can we distinguish or differentiate two phases via their low energy QP instead of the ground state? How to reveal the symmetry and internal structure of a QP from the entanglement spectrum?

In our previous work~\cite{you2020observing}, we scrutinized the QP entanglement in a 1D transverse-field Ising model with an integrability breaking perturbation. We demonstrated that the non-local QP states in the symmetry broken phase carry an additional `universal entanglement entropy' solely determined by the underlying broken symmetry. Such a universal entanglement entropy results from the non-locality of the QP which triggers a quantum coherence between globally distinct symmetry patterns. It follows that the non-locality of the QP in symmetry broken phases can be detected by computing the long-range mutual information (MI) between spatially separated qubits in the corresponding quantum state. This protocol provides a feasible way to detect non-local QPs in symmetry broken phases with long-range correlations.

In this paper, we extend our work to provide an entanglement protocol to detect various QPs in general 1D systems with short-range correlations. In accordance with the entanglement benchmark for the ground-state wave function in topological phases, we anticipate that the ES of the low energy QP can also differentiate and distinguish different SPT phases. In particular, we propose an `entanglement fragmentation' feature for QP states in free-fermion SPT phases with weak interactions, as a consequence of the hybridization between edge zero mode and bulk zero modes in the QP entanglement Hamiltonian. Under some circumstances, such an entanglement fragmentation pattern carries a universal particle-hole symmetry structure in each layer. This fragmentation feature provides a new guideline to identify different phase of matter via their unique QP excitations, and implies that the low energy QP state still inherits the topological structure of an SPT phase. We next extend our entanglement considerations to QP states in strongly interacting symmetry protected topological chains, and develop a conditional mutual information protocol to dissect the QPs in SPT chains. In particular, we show that the non-vanishing conditional mutual information indicates the emergence of non-local quantum orders in a ground state or in low energy QP states. While a non-local order parameter is operator dependent and hence hard to identify for a general wave function, the conditional mutual information is directly accessible from the reduced density matrix, and easy to measure in numerical simulations. In addition, we also propose a protocol based on the measurement induced long-range entanglement to identify distinct SPT QPs with potential experimental accessibility in cold atom setups~\cite{PhysRevLett.109.020504,Kaufman794,PhysRevLett.109.020505,PhysRevLett.120.050406,choo2018measurement,Brydges2019}. 

The structure of this paper is as follows: in Sect.~\ref{sec:pes} we discuss the correlation matrix spectra of QP states in non-interacting systems based on Peschel's approach~\cite{peschel2009reduced}, and illustrate the ideas by studying the paradigmatic SSH model. In Sect.~\ref{sec:kramers} we extend the discussion to interacting systems, and develop criteria for the degeneracy of the many-body QP ES based on the presence of reflection symmetry in combination with specific (e.g. $Z_2$) charges. In Sect.~\ref{sec:fragm} we then discuss the universal `fragmentation feature' in the ES of QP states in topological free-fermion systems. In Sect.~\ref{sec:mufs} we study the conditional MI and measurement-induced MI for the ground state and QP states of SPT phases. In particular, we show that both quantities provide feasible ways to probe SPT phases, not only via ground states but also via QP states.

\section{Correlation matrix approach for QP entanglement}\label{sec:pes}

In this section, we present a general framework to bridge the connection between the correlation matrix and the single-particle entanglement spectra for QP states in non-interacting systems, under a spatially symmetric bipartition, and study the SSH chain as a concrete example.
 
Previous work on the correspondence of the correlation matrix and entanglement Hamiltonian for free theory was carried out in Ref.~\cite{peschel2009reduced, turner2009band,fidkowski2010entanglement}. It was shown that the entanglement Hamiltonian for a gapped insulator is akin to a Hamiltonian matrix (with open boundary at the cut) where the states above/below the gap are flattened to energies $\infty$ and $-\infty$ respectively. Besides, if the insulator displays a topological band structure, its single-particle entanglement Hamiltonian contains an in-gap mode at zero energy akin to the edge mode in the physical Hamiltonian. This implies that, even without access to the full band structure, the ES can determine the band topology by scrutinizing the partial wave function of the ground state.

We begin our discussion by reviewing Peschel's approach~\cite{peschel2009reduced} for calculating the entanglement spectrum of a free-fermion system with trivial band topology by means of the correlation-matrix method.
For a non-interacting system where the Hamiltonian can be expressed in terms of fermion bilinears, the
reduced density matrix (RDM) with respect to region $A$ can be written as
\begin{equation}
\rho^A = e^{-H_A}/Z,
\end{equation}
where the constant $Z$ is written to impose normalization $\Tr \rho^A = 1$, and where $H_A$ is the so-called entanglement Hamiltonian. In this way the properties of the RDM can be reformulated in terms of thermodynamic properties of the entanglement Hamiltonian.
Then the correlation matrix for subsystem $A$ can be obtained as,
\begin{align} 
&C^A_{ij}=\langle c_i^{\dagger} c_j\rangle_{i,j \in A}=\Tr[ c_i^{\dagger} c_j \rho^A ]=\Tr[ c_i^{\dagger} c_j e^{-H_A}] /Z.
\end{align}
The entanglement Hamiltonian can be diagonalized in terms of new fermion operators $(\tilde{c}_{\alpha},\tilde{c}^{\dag}_{\alpha})$ which relate to the old ones by a canonical transformation
\begin{equation}
H_A = \sum_{\alpha} e_{\alpha} \tilde{c}^{\dag}_{\alpha} \tilde{c}^{\dag}_{\alpha},
\end{equation}
where $e_{\alpha}$ is the single-particle energy level of the spectrally flattened entanglement Hamiltonian.
 
Since the entanglement Hamiltonian represents a free-fermion theory, the expectation values of charge operators obey fermion statistics. We can then diagonalize the correlation matrix in terms of the new basis
\begin{align} 
&\langle \tilde{c}_{\alpha}^{\dagger} \tilde{c}_{\alpha}\rangle_{GS} =\Tr [\tilde{c}_{\alpha}^{\dagger} \tilde{c}_{\alpha}e^{-H_A} ]/Z
=\frac{1}{1+e^{e_{\alpha}}},
\end{align}
from which it follows that the eigenvalues of the correlation matrix $p_{\alpha}$ are related to the single-particle spectrum of $H_A$ as
\begin{align} \label{corre}
&e_{\alpha}=\ln [(1-p_{\alpha})/p_{\alpha}].
\end{align}
The many-body entanglement spectrum is the union of the quasienergies of all possible filled states, $E=\sum_{\alpha} n_{\alpha}e_{\alpha}$. If the energy level of $H_A$ resembles a gapped insulator, the many-body entanglement spectrum also exhibits a gap.

As the correlation matrix has eigenvalues $p_{\alpha}=0,1$ for an insulating ground state, the entanglement Hamiltonian $H_A$ has a finite gap for the bulk orbitals. For now, we focus solely on a trivial band insulator whose ground state entanglement Hamiltonian $H_A$ does not carry any in-gap state. We will return to the cases of topological insulators with in-gap states in Sect.~\ref{sec:fragm}.

\begin{figure}
\includegraphics[width=0.95\linewidth]{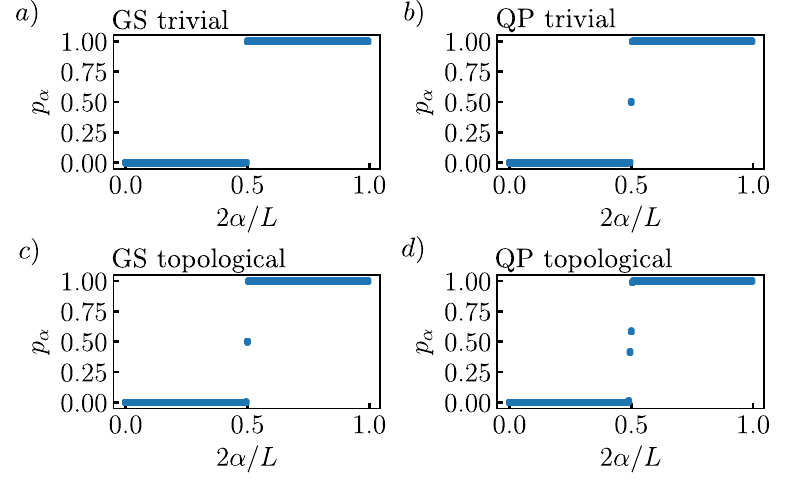}
\caption{The correlation matrix spectra of the GS and QP in the SSH chain. The simulations are performed on chains of $L=400$ sites, but we rescaled the number of states with the system size as the results are universal. a) For the GS in the trivial phase no in-gap states are present, half of the orbitals are filled and half are empty. b) There is an in-gap state for the QP in the trivial phase which is an extended bulk state reflecting the position uncertainty of the QP. c) There is also a in-gap state for the GS in the topological phase but this one is due to the protected boundary modes. d) For the QP in the topological phase there are two in-gap states, one reflecting the bulk mode and one the boundary mode. Note that when $\delta$ approaches zero (from the negative side, as we are in the topological phase) the two in-gap states move further apart, which represents further mixing of the bulk/boundary modes. They can only disappear into the empty or filled orbitals by crossing the phase transition point at $\delta =0$.} 
\label{fig:ssh_spes}
\end{figure}

\begin{figure}
\includegraphics[width=0.49\textwidth]{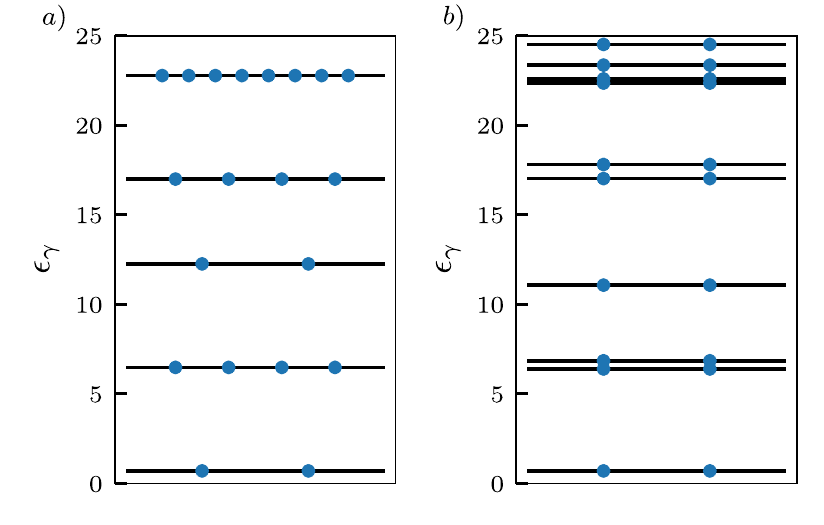}
\caption{The entanglement spectrum of the first excited state in the trivial SSH model on a chain with $L=20$ sites with $\delta=0.5$.  a) The non-interacting case with $U=0$.  b) The interacting case with $U=0.2$. Both ES exhibit two-fold degeneracy in all energy levels.} 
\label{fig:trivialssh}
\end{figure}

Now we add a QP from the conducting band to the ground state and make a center symmetric cut to obtain the sub-region $A$. We calculate the correlation matrix based on the QP state with momentum $k$,  

\begin{align} 
&C^{QP}_{ij}=\langle QP |c_i^{\dagger} c_j|QP \rangle_{i,j \in A}\nonumber\\
&=\langle GS| \frac{1}{\sqrt{2}} (a^k_A+a^k_B) |c_i^{\dagger} c_j| \frac{1}{\sqrt{2}} (a^{k\dagger}_A+a^{k\dagger}_B) |GS \rangle_{i,j \in A}
\end{align}
where $ a^{k\dagger}_A= \frac{1}{\sqrt{L/2}} \sum_{i \in A} e^{ikr_i}  a^{\dagger}_i$ is the operator which creates an excited state with momentum $k$ in the $A$ region. $a^{\dagger}_i$ represents the Wannier orbital in the upper-band, as this orbital might involve a linear combination of the degrees of freedom among several sites, there are some $a^{\dagger}_i$ operators near the cut sitting in both the $A$ and $B$ region. However, these operators can be ignored in the thermodynamic limit as they only carry a weight proportional to $\frac{1}{\sqrt{L}}$. Therefore,
\begin{align} 
&C^{QP}_{ij}=\langle QP |c_i^{\dagger} c_j|QP \rangle_{i,j \in A}\nonumber\\
\begin{split} &=\frac{1}{2}[ \langle  GS |c_i^{\dagger} c_j  (1-a^{k\dagger}_B a^k_B) | GS \rangle_{i,j \in A} \\ & \quad \quad \quad \quad \quad \quad  + \langle  GS|a^k_A c_i^{\dagger} c_j a^{k\dagger}_A| GS   \rangle_{i,j \in A}]
\end{split} \nonumber\\
&=\frac{1}{2}[\langle  GS |c_i^{\dagger} c_j  | GS \rangle_{i,j \in A}+ \langle GS |a^k_A c_i^{\dagger} c_j a^{k\dagger}_A| GS  \rangle_{i,j \in A} ], 
\end{align}
and if we diagonalize the correlation matrix in terms of the basis $\tilde{c}_{\alpha}$, we obtain the eigenvalues $p_{\alpha}$ of the correlation matrix 
\begin{align} 
&p_{\alpha}=\frac{1}{2}[\langle  GS |\tilde{c}_{\alpha}^{\dagger} \tilde{c}_{\alpha} | GS \rangle_{i,j \in A}+\langle GS |a^k_A\tilde{c}_{\alpha}^{\dagger} \tilde{c}_{\alpha} a^{k\dagger}_A| GS  \rangle_{i,j \in A}]\nonumber\\
\nonumber\\
&= \begin{cases}
&1 , ~~\alpha \in \text{filled orbitals} \nonumber\\
&0,  ~~\alpha \in \text{empty orbitals}  \nonumber\\
&1/2 , ~~\alpha \in \text{QP orbitals.} 
 \end{cases}  
\end{align}
The eigenvalues of the correlation matrix with respect to the QP state have three types of eigenstates. For filled (or empty) orbitals, it remains $1$ (or $0$) like the correlation matrix of the ground state. The additional QP orbital generates an eigenvalue $p_{\alpha}=1/2$ independent of the QP momentum. This $p_{\alpha}=1/2$ mode can be understood as the position uncertainty of the QP: with a probability $1/2$ the QP is in region $A$/$B$ and this thus creates an additional contribution to the entropy. Based on the correspondence in Eq.~\eqref{corre}, this additional $p_{\alpha}=1/2$ mode is equivalent to the statement that the entanglement Hamiltonian $H^{QP}_A$ contains an exact zero mode $e_{\alpha}=0$ inside the gap for the QP in a trivial system (we will illustrate this in Fig.~\ref{fig:ssh_spes}$b$ for the trivial SSH chain). In particular, as the $p_{\alpha}=1/2$ mode signifies the QP orbital extended in region $A$, the in-gap mode in the entanglement Hamiltonian denotes an extended bulk state in $H^{QP}_A$ so the entanglement Hamiltonian for the QP resembles a metallic system with one conducting state in the bulk. In the presence of such in-gap zero mode, the many-body entanglement spectrum $\{ E=\sum_{\alpha} n_{\alpha}e_{\alpha}\}$ exhibits a two-fold degeneracy for all energy levels as a consequence of the empty/filled in-gap state (see also Fig.~\ref{fig:trivialssh}). 

\begin{figure}
\includegraphics[width=0.95\linewidth]{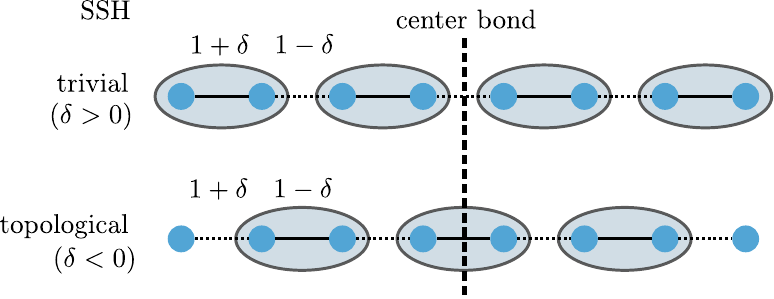}
\caption{A sketch of the SSH model in our setup: we consider chains of length $L=4m$ (with $m$ integer), such that we have an even number of unit cells in the trivial phase and an odd number in the topological phase. In the trivial phase the center bond is weak, while in the topological phase it is strong.  } 
\label{fig:ssh_sketch}
\end{figure}

We illustrate the theorem by simulating the 1D Su-Schrieffer-Heeger (SSH) model~\cite{ssh1979} 
  \begin{align} \label{eq:SSH}
  H= \sum_{i=0}^{L-1} ( (1+\delta)c^{\dagger}_{2i} c_{2i+1} &+ (1-\delta)c^{\dagger}_{2i+1} c_{2i+2} ) \nonumber \\
  &+ V(n_0 + n_{L-1}),
\end{align}   
with $V>0$. For $\delta <0$ this model has weak bonds on the edges and is topological, while for $\delta>0$ it is trivial and resembles an atomic insulator. We will always simulate chains of length $L=4m$ (with $m$ integer) such that the center bond is always weak (strong) in the trivial (topological) phase, and such that there is an even number ($2m$) of two-site unit cells in the trivial phase. These properties are summarized in Fig.~\ref{fig:ssh_sketch}. We additionally choose to add a boundary field $V$ in order to lift the four-fold degeneracy of the ground state in the topological regime; this does not change the physics. Here, we will demonstrate the above statements for the trivial case, and delay the discussion of the topological phase to Sect.~\ref{sec:fragm}. In the trivial phase the GS is located at half filling such that the correlation matrix spectrum will just consists of zeros and ones, as shown in Fig.~\ref{fig:ssh_spes}$a$. We then create a QP state by adding an additional orbital with zero-momentum from the empty band. The resultant correlation matrix spectrum is shown in Fig.~\ref{fig:ssh_spes}$b$. The in-gap metallic state is clearly visible, and indicates the QP position uncertainty as outlined above. For that reason the many-body entanglement spectrum is also two-fold degenerate for all levels as shown in Fig.~\ref{fig:trivialssh}$a$.

Our current argument relies on the circumstance that the system is non-interacting, and that the QP is based on a band insulator with trivial topology. For QP states in topological insulators or superconductors, the band topology adds more heterogeneity to their entanglement structure. In particular, the entanglement Hamiltonian of the ground state already contains an in-gap state which resembles the gapless edge (Fig.~\ref{fig:ssh_spes}$c$). In the forthcoming discussion in Sec.~\ref{sec:fragm}, we will discuss a universal fragmentation structure of the QP ES in topological band theory as a consequence of the level mixing between the edge and QP state.

Besides, the stability of the exact zero modes in the entanglement Hamiltonian, which leads to a two-fold degeneracy in the QP ES is only evident for free-fermion systems with trivial band topology. In the presence of interactions, it remains unclear whether the doubling in the ES still survives. In the next section, we proceed to explore the ES in strongly-interacting systems.

\section{Kramers theorem for QP entanglement Hamiltonian} \label{sec:kramers}

In our discussion in the previous section, we mentioned the two-fold degeneracy of the ES of the QP in a trivial SSH chain under a symmetric cut. Such a doubling in the spectrum originates from the QP position uncertainty for being either inside or outside of the cut. In a free-fermion system with trivial topology, this doubling of the ES is manifested by the exact zero modes of the single-particle entanglement Hamiltonian. When it comes to the interacting system, it remains unclear whether the degeneracy of the QP ES would persist as the entanglement Hamiltonian cannot be reduced to a single particle operator.

In this section, we develop a universal criterion for the degeneracy of the QP ES based on a Kramers theorem. The basic idea can be traced back to our previous work~\cite{you2020observing} where we examined the symmetry representation of the QP reduced density matrix. For instance, if the reduced density matrix has a projective representation under a symmetry $G$, the entanglement spectrum must display a degeneracy for all energy levels. We will study some concrete cases in the remainder of this section.

\subsection{$Z_2$ and reflection}\label{sec:sym}
To set the stage, we reiterate the QP ES degeneracy theorem in the 1D Ising paramagnetic phase of Ref.~\cite{you2020observing}. We define the model with an integrability-breaking term
 \begin{align} 
  &H= \sum_i J_z \sigma^z_i\sigma^z_{i+1}+h_x \sigma^x_i+J_x \sigma^x_i\sigma^x_{i+1},
  \end{align}
 and consider the regime where $h_x>J_z \gg J_x$ such that the ground state is polarized in the $x$ direction and the QP---which we will refer to as a $Z_2$ magnon---is generated by a local spin flip operator $Q_{i}=\sigma^z_i$ which creates a $Z_2$ charge measured by the parity operator $P=\prod_i \sigma^x_i$. Away from the extreme paramagnetic limit $Q_i$ remains an odd operator under $Z_2$ but now spreads out over a correlation length $\xi$. With periodic boundary conditions, the system is translationally invariant and the low energy magnon states are momentum eigenstates of the form \cite{castro2018entanglement}, 
\begin{align} 
  &|\psi^{1M}\rangle_k = \sum_i e^{i k r_i}Q_i |\mathrm{GS} \rangle
  \end{align}
where we have ignored the subleading $k$ dependence of the operator $Q_i$ itself. 

We now consider the entanglement spectrum of the magnon at $k=0$, defined as $\epsilon_{\gamma} = -\ln \lambda^2_{\gamma}$, which is obtained from a Schmidt decomposition 
\begin{equation}
|\psi^{1M}\rangle= \sum_{\gamma} \lambda_{\gamma}|\gamma\rangle_A |\gamma\rangle_B.
\end{equation}
Note that $\lambda_{\gamma}^2$ are the eigenvalues of the reduced density matrices $\rho_A$ and $\rho_B$ for the two subsystems. For a reflection symmetric cut, the entanglement spectrum exhibits an {\it exact} two-fold degeneracy. As is pointed out in Ref.~\cite{you2020observing}, this exact degeneracy of the Schmidt values arises from a combination of the reflection symmetry $R|\psi^{1M} \rangle=|\psi^{1M} \rangle$ and the non-trivial $Z_2$ charge of the magnon $P|\psi^{1M} \rangle=-|\psi^{1M} \rangle$. 

\begin{figure}
\includegraphics[width=0.45\textwidth]{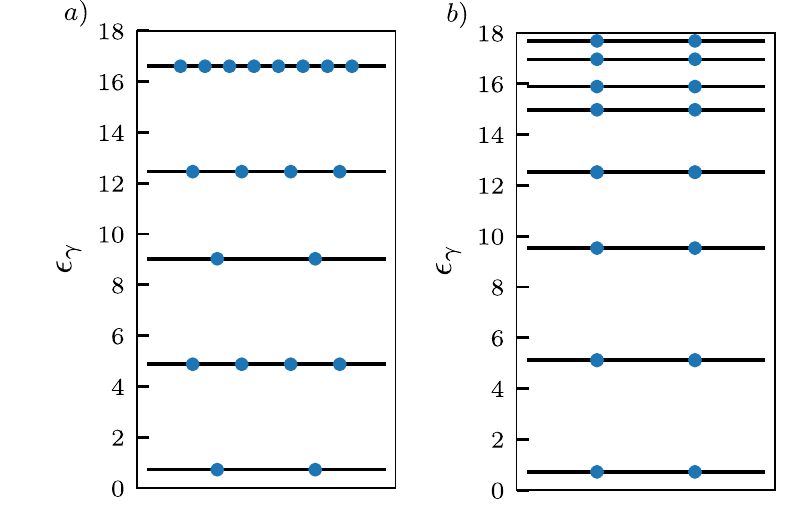}
\caption{The entanglement spectrum of the first excited state in the PM Ising model (18 sites, $h_x=1.5$).  a) The non-interacting case with $J_x=0$.  b) The interacting case with $J_x=0.2$. Both ES exhibit two-fold degeneracy in all energy levels.} 
\label{z2case}
\end{figure}

Assume that the entanglement spectrum contains a non-degenerate eigenvalue $\lambda_{\gamma}$.
Then the fact that the state $|\psi^{1M}\rangle$ is symmetric under $R$ whence $|\gamma\rangle_A =R|\gamma\rangle_B$ up to a $U(1)$ phase.
However, as the QP state carries an odd charge parity we have that $P|\psi^{1M} \rangle=-|\psi^{1M} \rangle$, hence $|\gamma\rangle_A$ and $|\gamma\rangle_B$ must have opposite parity eigenvalues. This leads to a contradiction with the requirement $|\gamma\rangle_A =R|\gamma\rangle_B$ as $R$ does not change the charge.
Consequently, all Schmidt values have to be degenerate. This is illustrated in Fig.~\ref{z2case} for both the non-interacting ($J_x= 0$) and interacting ($J_x \neq 0$) case.

  \subsection{$Z_2$ charge and translation}
  The arguments in our previous section can be generalized to magnon QP states with $k \ne 0$ where instead of reflection we can use translations over half the system size $T_x=e^{i k L/2}$ and impose periodic boundary conditions. We now consider the magnon entanglement spectrum of the QP at $k$, defined as $\epsilon_{\gamma} = -\ln \lambda^2_{\gamma}$, which is obtained from a Schmidt decomposition, 
\begin{equation}
|\psi^{1M}\rangle_k= \sum_{\gamma} \lambda_{\gamma}|\gamma\rangle_A |\gamma\rangle_B.
\end{equation}
Again, we begin by assuming that the ES contains a non-degenerate eigenvalue $\lambda_{\gamma'}$. The state $|\psi^{1M}\rangle$ being symmetric under $T_x$ implies $|\gamma'\rangle_A =R|\gamma'\rangle_B$ up to a phase factor. So the pattern $\lambda_{\gamma'}|\gamma'\rangle_A |\gamma'\rangle_B$ carries even charge parity. This leads to a contradiction with the fact that the QP $|\psi^{1M}\rangle_k$ carries odd charge parity. Consequently, all Schmidt values have to be degenerate.

 \subsection{$Z_3$ charge}
  
In this section, we generalize our study of QP entanglement to $Z_M$ Ising models whose QPs carry $Z_M$ charges. We begin by examining the QP ES in the in the paramagnetic phase of the $Z_3$ clock model~\cite{fendley2012parafermionic,ostlund1981incommensurate} 
  
  \begin{align} 
  &H= \sum_i ~\sigma^{\dagger}_i\sigma_{i+1}+h \tau_i,\nonumber\\
  &\tau_i \sigma_j=e^{i2\pi/3}\delta_{ij}\sigma_j \tau_i \nonumber\\
  & \tau^3_i = \sigma_i^3 = I.
  \end{align}

The QP state with zero momentum carries charge 1 (modulo 3), and is both reflection and $Z_3$ symmetric. However, neither of these symmetries, individually or in combination, constitutes a projective symmetry for the RDM of the half chain. In particular, consider a center symmetric cut for which a QP wave function has the Schmidt decomposition,
\begin{equation}
|\psi^{1M}\rangle= \sum_{\gamma} \lambda_{\gamma}|\gamma\rangle_A |\gamma\rangle_B.
\end{equation}
This decomposition contains identical configurations $|\gamma\rangle_A =|\gamma\rangle_B$, each of which carries charge 2 since they add up to charge unity modulo 3. Such configurations can be associated with a unique energy level in the ES. Hence, the QP ES in $Z_3$ paramagnetic chain does not exhibit a generic degeneracy. 
 The same conclusion applies for other QP states in $Z_{2m+1}$ clock model.

 Likewise, for $Z_{2m}$ Ising PM phase, the QP wavefunction carries an odd charge so the reduced density matrix for each half-chain renders a symmetric partition of even and odd charge patterns. This again enforces a two-level degeneracy in the entanglement spectrum.

\subsection{U(1) charge and reflection}

 The treatment of QPs carrying a $U(1)$ charge is similar to the $Z_2$ case, as $Z_2$ is a subgroup of $U(1)$, so the argument in Sec.~\ref{sec:sym} applies.
  Based on this argument, we begin with a zero momentum QP state in a trivial SSH chain and obtain its entanglement spectrum with respect to the center cut. As the zero momentum QP state carries an odd charge and is reflection symmetric, we expect a robust two-fold degeneracy in the entanglement spectrum. We verify this conjecture by adding interactions (that do not break the $U(1)$ symmetry) to the SSH chain~\eqref{eq:SSH}, such that the model becomes  
    \begin{align}\label{eq:SSH_int}
  H= \sum_{i=0}^{L-1} ( (1+\delta)&c^{\dagger}_{2i} c_{2i+1} + (1-\delta)c^{\dagger}_{2i+1} c_{2i+2} ) \nonumber \\
  &+U n_i n_{i+1} + V(n_0 + n_{L-1}).
\end{align}  
where $n$ is the onsite charge density away from half-filling, and $V>0$.
We construct the QP state in the trivial phase for a chain of 20 sites with $U=0.2$, and obtain the entanglement spectrum from a center cut. As long as the reflection symmetry and $U(1)$ symmetry are preserved in the QP state, the Kramers theorem still applies regardless of the interaction strength and the two-fold degeneracy remains robust as shown in Fig.~\ref{fig:trivialssh}$b$.

It is worth recalling that for the SSH chain in the topological phase, the ground state wave function contains an odd charge and its ES is fully degenerate and protected by reflection and $U(1)$ symmetry. However, when it comes to the QP state, the total charge is even. Therefore, strictly speaking, we do not expect any protected degeneracy for the QP ES. Nevertheless, such QP state in the topological phase exhibits a `fragmentation feature' which will be elaborated in the following section.

\section{Universal Entanglement fragmentation, a fingerprint for SPT quasiparticles} \label{sec:fragm}

The relation between 1D SPT phases and their entanglement properties has been widely explored in Refs.~\cite{Li-2008,Peschel_2009,pollmann2012symmetry,PhysRevB.84.195103,PhysRevLett.105.115501,PhysRevLett.104.130502,chandran2014universal,PhysRevLett.110.236801,Pollmann2010,Pollmann2012-lv,pollmann2012symmetry,Turner2011-zi,fidkowski2010entanglement}. Remarkably, it has been demonstrated that 1D SPTs can be characterized by their ground-state ES~\cite{Pollmann2010,Pollmann2012-lv,pollmann2012symmetry,Turner2011-zi}. More precisely, it was shown that the irreducible representation of the projective symmetry in the matrix product state (MPS) determines the degeneracy of the GS ES, and hence can be treated as a fingerprint of SPT states. Similar ideas has been applied to a wide variety of SPT phases in higher dimensions with higher-order topology~\cite{you2020higher}.

The universal entanglement behavior of QP states in SPT phases remains unclear. In particular, despite the fact that the SPT GS captures salient features of the underlying topology, it is not known whether these features are inherited by the QP excited states at low energy. A more general understanding of the entanglement properties of QP states in SPT phases is therefore left open. In this Section, we try to answer this question by proposing a universal `entanglement fragmentation' feature for SPT quasiparticles. Such a feature indicates that the low energy QP states in a non-interacting SPT phase still retain some of the topological features of the ground state. Based on this observation, one can still identify different phases via the entanglement in the QP states.

\subsection{QP of topological SSH chain}

 Let us begin with the example of the topological Su-Schrieffer-Heeger (SSH) chain protected by $U(1)$ and reflection symmetry ($R$), whose ground state ES exhibits a two-fold degeneracy under a center symmetric cut. Assume the GS has the Schmidt decomposition,
 \begin{align} 
 &|GS\rangle = \sum_{\gamma} \lambda_{\gamma}|\gamma\rangle_A |\gamma\rangle_B.
  \end{align}
  Due to the strong dimer bond between the cut, the charge difference $q_A-q_B$ between left and right half of the wave function is always an odd number, as the total number of particles in the ground state is odd. Following the discussion in Sec.~\ref{sec:sym}, the GS entanglement spectrum $\epsilon_{\gamma} = -\ln \lambda_{\gamma}^2$ must exhibit two-fold degeneracy.
  This degeneracy also has a simple explanation in terms of the single-particle entanglement Hamiltonian introduced in Sec.~\ref{sec:pes}: the existence of the entanglement in-gap mode, shown in Fig.~\ref{fig:ssh_spes}$c$, reflects the character of the topologically-protected mode at the boundary of the system~\cite{Turner2011-zi,turner2009band,fidkowski2010entanglement}.

Now we consider the QP excited state with zero momentum. Adding such QP to the filled topological band still preserves the $P$ symmetry, and its position uncertainty should give rise to another in-gap zero mode, as we have showed in Sec.~\ref{sec:pes}, which is an extended bulk state. Subsequently, the QP's entanglement Hamiltonian $H_A^{QP}$, in the zero-correlation length limit (i.e. the fully dimerized limit on the topological side), contains two zero modes $p_{\alpha} = 1/2$ labeled by $f$ and $c$ which corresponds to the bulk and edge in-gap mode. The entanglement Hamiltonian can be viewed as a metallic system with one conducting state in bulk in addition to one edge mode. 

\begin{figure}
\includegraphics[width=0.95\linewidth]{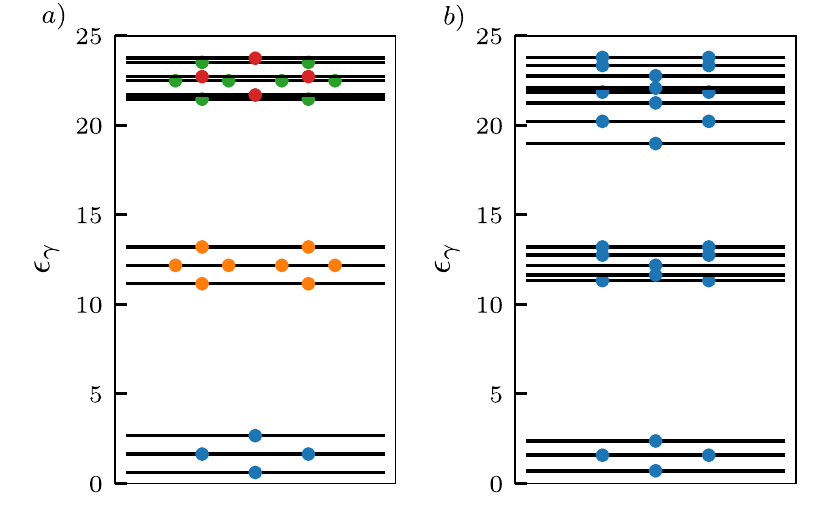}
\caption{The entanglement spectrum of the first excited state in the topological SSH model on a chain with $L=20$ sites and $\delta=-0.5$. a) The non-interacting case with $U=0$. The ES exhibit a clear fragmentation pattern with each fragment being particle-hole symmetric. The colors are added to make the fragmentation pattern clearly visible. b) The interacting case with $U=0.2$. The fragmentation pattern is clear for the low energy patterns but not for the higher ones. The particle-hole symmetry within each fragment is broken.} \label{fig:topossh}
\end{figure}

Now we consider the case with finite correlations. This is like turning on a symmetry allowed coupling $f^{\dagger}c+h.c.$ which hybridizes these bulk and edge zero modes, and thus lifts the degeneracy and delocalizes the edge zero mode. In particular, terms like $f^{\dagger}f, c^{\dagger}c$ are still absent from the entanglement Hamiltonian as they would break the reflection symmetry. Consequently, the degeneracy lifting between two zero modes is particle-hole symmetric with energy levels $\pm \epsilon$ as shown in Fig.~\ref{fig:ssh_spes}$d$. In the many-body ES, these splitting modes prompt an `entanglement fragmentation' pattern with each fragment containing four eigenvalues $E_a+\epsilon,E_a-\epsilon,E_a,E_a$ as shown in Fig.~\ref{fig:topossh}. The states $E_a \pm \epsilon$ come from the situation where either the $\epsilon$ or $-\epsilon$ mode is filled. The two degenerate $E_a$ states originate from the situation where the two modes are either completely filled or completely empty.

For the non-interacting theory, this entanglement fragmentation feature has some universal properties that are independent of the microscopic Hamiltonian. 1) Each layer in the fragmented entanglement spectrum has three energy levels ($E_a,E_a \pm \epsilon$), with the middle one ($E_a$) being $2n$-fold degenerate. 2) The top and bottom energy level are $n$-fold degenerate, and are related to the middle one via particle-hole symmetry. 
Based on this argument, we found an immediate distinction between QPs from distinct phases in the SSH chain: in the topological phase the QP carries a universal entanglement fragmentation pattern, while in the trivial phase the QP exhibits a 2-fold degeneracy due to Kramers theorem. Because of its inherent robustness, such a unique QP entanglement fragmentation pattern is expected for most non-interacting SPT systems.

This entanglement fragmentation feature has profound implications. Beyond analyzing the usual ground state patterns, one could also investigate the entanglement spectrum of the low-energy QP states to distinguish between different phases. In particular, the universal fragmentation pattern in the low-energy QP of the topological SSH chain, implies that some topological structure is still present.

Our current analysis is based on the non-interacting SSH chain. In the free-fermion limit, the coupling between two in-gap modes is always particle-hole symmetric so their level mixing also inherits a particle-hole symmetric fragmentation pattern. 
The situation in the interacting case is far from our previous scope. Henceforward, we will focus on a more general understanding of the QP entanglement fragmentation in the presence of strong interactions.

We start this argument by detailing the level mixing between two zero modes.
In the zero correlation length limit, the two modes are decoupled in the thermodynamic limit. The reflection symmetry $R_x$ sends
\begin{align} 
f \rightarrow f^{\dagger},~c\rightarrow c^{\dagger},
  \end{align}
  which acts like a particle-hole symmetry for the zero mode and hence prohibits a potential term to lift the degeneracy.
  For the non-interacting theory, one can turn on the coupling $f^{\dagger}c+h.c.$ to induce hybridization between these bulk and edge zero modes and thus lift the degeneracy with energy $\pm \epsilon$. In particular, as there is only a single bulk zero-energy mode, their energy mixing scales with system size in a power-law manner $\epsilon \sim 1/L$. Hence, in free theory, the reflection and $U(1)$ symmetry ensure that the energy splitting in each entanglement fragmentation is particle-hole symmetric.

Now we look into the interacting case. One can turn on a symmetry invariant interaction
\begin{align} 
H_{int}= U (f^{\dagger}f-1/2)(c^{\dagger}c-1/2),
  \end{align}
which creates a gap between different charge parity sectors such that all filled or all empty states $n_f=n_e$ have larger energy. This would break the particle-hole symmetric pattern in the entanglement fragmentation and thus mix different fragments in the ES. As a result, for each fragment, the middle energy level with two-fold degeneracy is no longer symmetric with respect to the top and bottom energy level with odd charge parity. Thus, the particle-hole symmetric fragmentation pattern is lost in the presence of interactions as shown in Fig.~\ref{fig:topossh}. 

\subsection{QP of topological SSH model with periodic boundary conditions}
\begin{figure} 
\includegraphics[width=0.8\linewidth]{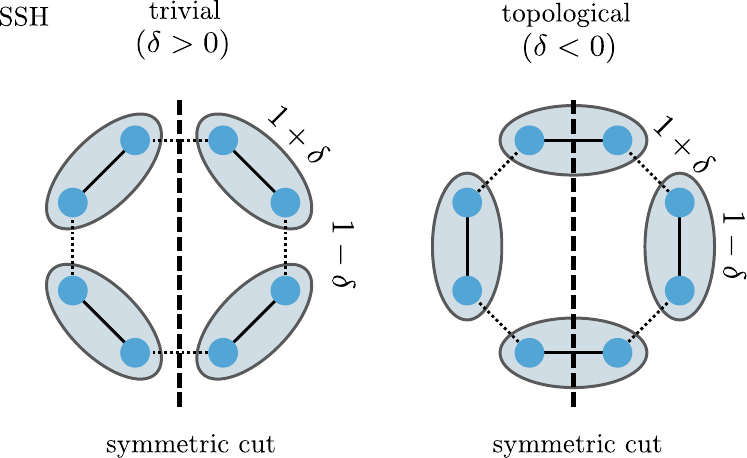}
\caption{Sketch of the SSH model with PBC in our setup, we choose $L=4m$ sites (with $m$ integer) such that the trivial (topological) phase is characterized by the cutting of weak (strong) bonds.} 
\label{fig:ssh_pbc_sketch}
\end{figure}

\begin{figure} 
\includegraphics[width=0.99\linewidth]{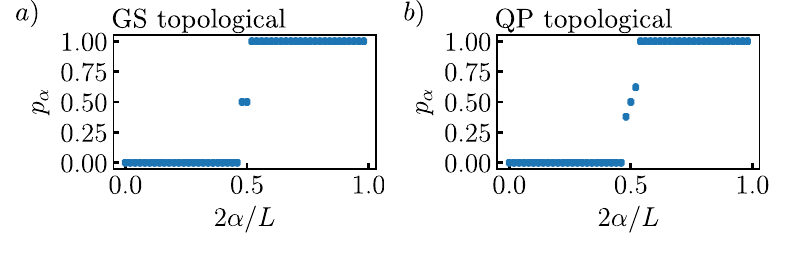}
\caption{The correlation matrix spectra for the topological SSH model with PBC. a) The spectrum of the GS is characterized by two states $p_{\alpha}=1/2$ because of the two boundary modes -- one near each cut. b) For the QP in the topological phase, two in-gap edge states are hybridized via the zero energy bulk state in a particle-hole symmetric way.} 
\label{fig:ssh_pbc_spes}
\end{figure}

Now we consider the SSH model~\eqref{eq:SSH} with periodic boundary conditions (PBC) and $V=0$ in the topological regime. As sketched in Fig.~\ref{fig:ssh_pbc_sketch}, this means that two strong bonds are cut in a symmetric bipartition of the system.
For the ground state, we will encounter an entanglement spectrum with four-fold degeneracy due to the two zero modes $c^{\dagger}, d^{\dagger}$ in the entanglement Hamiltonian localized near each cut, see Fig.~\ref{fig:ssh_pbc_spes}$a$.
The quasiparticle state adds an additional extended zero mode $f^{\dagger}$ to the entanglement Hamiltonian and makes the entanglement Hamiltonian like that of a metallic system with one conducting state in the bulk in addition to two edge modes. We expect these two edge modes to hybridize via the bulk conducting state and thus lift the degeneracy away from the zero correlation length limit, as shown in Fig.~\ref{fig:ssh_pbc_spes}$b$.

\begin{figure}
\includegraphics[width=0.95\linewidth]{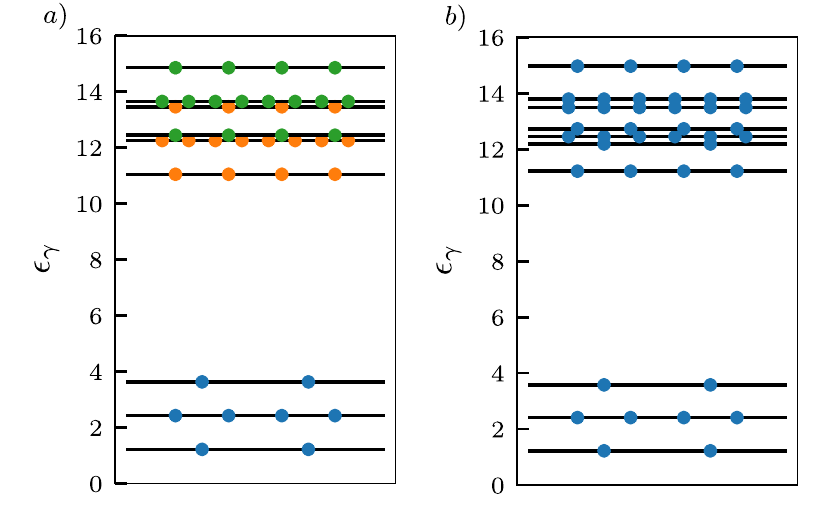}
\caption{Many-body entanglement spectra for the topological SSH model with PBC for a ring with $L=20$ sites and $\delta=-0.5$. a) Non-interacting case $U=0$. All quasienergy levels are two-fold degenerate due to one in-gap zero mode. The entire ES exhibits a clear fragmentation pattern with each fragment being particle-hole symmetric. 
b) Interacting case $U=0.1$. The fragmentation pattern is destroyed for all but the lowest levels, but the two-fold degeneracy remains robust.} 
 \label{fig:ssh_pbc_es}
\end{figure}

  We can turn on the reflection invariant coupling $f^{\dagger}c+f^{\dagger}d+h.c$ to hybridize the three zero modes into energies $0$ and $\pm \epsilon$ which is still a particle-hole symmetric spectrum. The zero energy mode again implies that the entanglement spectrum should have a two-fold degeneracy, so that the fragmentation structure is doubled compared to the single cut with open boundary. This doubling is shown in Fig.~\ref{fig:ssh_pbc_es}. The Kramers theorem still applies as the QP state has odd charge parity. So this two-fold degeneracy is robust against any $U(1)$ preserving interaction, as depicted in Fig.~\ref{fig:ssh_pbc_es} for the model~\eqref{eq:SSH_int} with $U=0.1$ and $V=0$, but the fragmentation patterns are mostly lost.

\subsection{$\mathcal{T}$ symmetry}
We briefly comment on an extension of our considerations to an interacting bosonic system with $\mathcal{T}$ symmetry. 
As a concrete example, we consider the spin-1 Haldane chain~\cite{affleck1988valence,Haldane1983-ya} with a half-integer spin excitation at the boundary.
The ES of the ground state with PBC has a four-fold degeneracy due to the projective symmetry at each cut (up to exponentially small finite size effects).
These two projective symmetries, represented by a half-spin degree of freedom, cannot be hybridized via a finite-size local unitary circuit. 

Now we add a triplon excitation to the ground state. 
We focus on the condition where the QP is a triplet excitation between the spin-1/2 singlet on the bond  which still respects the $\mathcal{T}$ and easy-plane $U(1)$ symmetry for $S_z$. 
Despite the fact that the QP itself is a Kramers singlet, it is composed of two spin-1/2 from the nearby sites and thus it renders a projective representation per site.
Thus, the two `edge states' of the entanglement Hamiltonian are now coupled via this extended QP state in bulk and the degeneracy is lifted.
This can formally also be understood using an MPS:
The tensors $M^m$ of the MPS representation of an SPT state $|\mathrm{GS}\rangle$ transform projectively such that $M^m \rightarrow U^{\dag}_{\mathcal{T}} M^m U_{\mathcal{T}}$ with $U_{\mathcal{T}}U_{\mathcal{T}}^*=-1$.
The projective representation implies degeneracies of the ES.
The QP state $|\mathrm{QP}\rangle = \sum_r e^{ikr}O_r |\mathrm{GS}\rangle$, with $O$ creating locally a triplon excitation, however mixes linear and projective representations and thus the degeneracies in the ES are lifted.

\section{Conditional Mutual information for quasiparticles in SPT chains} \label{sec:mufs}

Despite the lack of local order or long-range correlations, the ground states of 1D SPT phases bear a `hidden non-local order' furnished by local entanglement patterns. Such `hidden non-local order' can be quantitatively defined and detected by measuring string-order parameters in the ground state wave function. While string-order parameters do exist for all SPT phases, these can be rather involved to measure for more complex phases \cite{PhysRevLett.109.050402, PhysRevB.86.125441}.
In this section, we will provide a quantum information perspective 
to leverage the relation between string order and conditional mutual information. In particular, we demonstrate that the non-local string order can be comprehended as the conditional long-range mutual information~\cite{jian2015long} of the wave function where a measurement of qubits at the two ends of a string reduces the information entropy of the total qubits on the string. When it comes to the QP state of the SPT phase, albeit its vanishing string order, the conditional mutual information is still nonzero and hence provides a feasible way to detect SPTs via QP entanglement.

\subsection{GS properties}\label{sec:gs_mi}  
\subsubsection{Mutual information and conditional mutual information}
\begin{figure}
\includegraphics[width=0.45\textwidth]{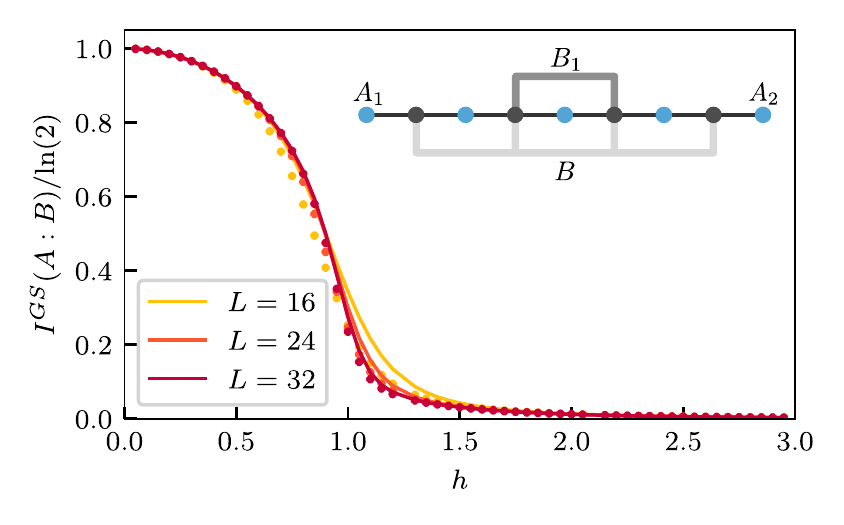}
\caption{The MI in the GS of the cluster model between the grey spins in the middle ($B$ area) and the blue spins ($A = A_1\cup A_2$) at the boundaries of the string. In the SPT phase $h_x <1$ it is non-vanishing for the GS. The dotted data points are the conditional MIs~\eqref{eq:cond_MI}, where we subdivided the region $B$ such that $B_2 = B \setminus B_1$ is the region the closest to the $A$ spins. 
There is only a small difference between the MI and conditional  MI  near  the  transition  as  the correlations are still dominantly local.} 
\label{fig:tfcluster_mi_gs}
\end{figure}

We start our discussion by considering a simple model of an SPT phase in 1D, namely the transverse-field cluster model~\cite{Suzuki1971,Briegel2001,WenBook2015}
\begin{equation}\label{eq:tf_cluster}
H = K \sum_i \sigma^z_{i-1} \sigma^x_{i} \sigma^z_{i+1} + h \sum_i \sigma^x_{i}.
\end{equation}
The SPT phase is protected by a $Z_2 \times Z_2$ symmetry generated by $Z_2^a = \prod_i \sigma^x_{2i+1}$ and $Z_2^b = \prod_i \sigma^x_{2i}$, and the system undergoes a phase transition from SPT to trivial at $h=1$. In the presence of an open boundary, each edge gives rise to a two-fold degeneracy due to the projective symmetry generated by the two Ising ($Z_2$) symmetries at the boundary. It is noteworthy to mention that the ground state contains a non-vanishing string order~\cite{denNijsRommelse}, 
  \begin{align} 
  \sigma^z_{2i}(\prod^{m-1}_{j=0}\sigma^x_{2i+2j+1})\sigma^z_{2i+2m}=1. 
    \end{align}
  This non-vanishing string order is a direct consequence of the decorated domain wall condensate because the total $Z_2^a$ charge living on the odd sites of the string is locked to the two spin configuration at the edge of the string (on the even sites). When the spins at the $2i,2i+2m$ sites are in the $|00\rangle,|11\rangle$ pattern, the total $Z_2^a$ charge between them is even.
 Such non-vanishing hidden order can be detected via the mutual information between the two spins $(A)$ at positions $2i$ and $2i+2m$ living at the ends of the string, and the odd site spins $(B)$ living inside the string as shown in Fig.~\ref{fig:tfcluster_mi_gs}
   \begin{align} 
I(A:B)=S(B)+S(A)- S(A\cup B)\neq 0.
  \end{align}
  Due to the decorated domain wall condensate structure, the spin pattern in the ground state is strongly fluctuating. Hence, the entanglement entropy of region $A$ converges to $S(A)=2\ln (2)$ for large $m$. Likewise, for the $m$ spins living on the odd sites of the string, referred to as region $B$, is $S(B)=m\ln(2)$ as each spin can have two configurations. However, if we compute the entropy $S(A\cup B)$ of the $m+2$ spins, it is $\ln(2)$ smaller than the entropy addition  $S(A)+S(B)$. The reason is obvious: by fixing the pattern of the spin at the end of the string, the total $Z_2^a$ charge on the odd sites is fixed and the entropy is reduced.

We performed density-matrix renomalization group (DMRG)~\cite{White1992} simulations to calculate the MI for the model~\eqref{eq:tf_cluster} with PBC, such that the ground state is non-degenerate. In our simulations we choose the (even) sites as $A_1=\frac{L}{4}, A_2=\frac{3L}{4}$ and use system sizes $L$ divisible by eight. The results are shown in Fig.~\ref{fig:tfcluster_mi_gs}, where we indeed observe a non-vanishing MI in the topological phase that approaches $\ln(2)$ for the fixpoint $h=0$. Note that we need to construct the reduced density matrix for the subsystem $A\cup B$, and are therefore limited to rather small subsystems of the order of $12$ sites. 

Alternatively, we could consider R\'{e}nyi versions of the entropies in the MIs
\begin{equation}
S^{\alpha} = \frac{1}{1-\alpha} \log \Tr (\rho_A^{\alpha})    
\end{equation}
and use, e.g., the replica trick to compute those. In that way larger system sizes can be accessed when using MPS. In addition, there exist efficient experimental protocols based on randomized measurements to obtain $S^{2}$  ~\cite{PhysRevLett.109.020504,Kaufman794,PhysRevLett.109.020505,PhysRevLett.120.050406,choo2018measurement,Brydges2019}.

Although one expects that a non-vanishing string order creates a non-vanishing MI $I(A:B)$ between the `edge' and the `bulk' of the string, a non-vanishing MI itself cannot be treated as a solid manifestation of the string order. Since some of the sites in $B$ region are adjacent to $A$, the local entanglement between adjacent sites can add some trivial contribution to the mutual information. In order to exclude such trivial contributions, we divide $B$ into two regions: the center region $B_1$ far from the string end, and the two parts $B_2$ within the correlation length of the string boundary.

We then define the conditional MI~\cite{ben2020disentangling},
  \begin{align} \label{eq:cond_MI}
I(A:B_1 | B_2)=I(A:B_1\cup B_2)-I(A:B_2).
  \end{align}
This conditional MI computes the `hidden correlation' between the bulk and boundary of the string while excluding the `trivial' MI induced by local entanglement. This is also shown in Fig.~\ref{fig:tfcluster_mi_gs} by the dots, where we took the region $B_2$ to be the spins directly neighboring the $A$ spins at positions $\frac{L}{4}+1$ and $\frac{3L}{4}-1$. Only near the phase transition there is an offset between the data, illustrating that the non-vanishing mutual information is indeed due to the string order.

\subsubsection{Measurement induced non-local entanglement}
\begin{figure}
\includegraphics[width=0.45\textwidth]{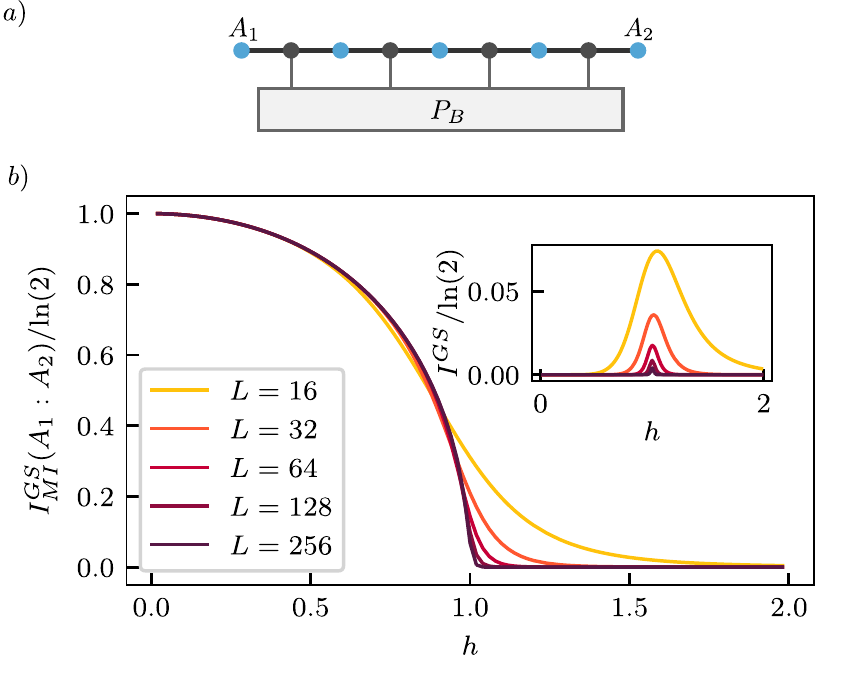}
\caption{a) Sketch of the projection of the odd sites in between $A_1$ and $A_2$ to a sector with definite (e.g. even) parity. b) The measurement induced MI in the ground state of the cluster model $P_B|GS\rangle$. The inset shows the data without the projection of the region $B$ to the even sector. In this case the distant spins in the $A$ region do not carry any mutual information in the thermodynamic limit, as expected. } 
\label{fig:mimi_gs}
\end{figure}

An alternative understanding of the MI in the GS can be obtained by introducing the concept of measurement induced non-local entanglement, where a charge measurement of the $B$ region $P_B|GS\rangle$, with
\begin{equation}
P_B = 1+\prod_{i \in B} \sigma^x_{i},
\end{equation}
triggers a dramatic non-local entanglement among the two spins in the $A$ region. A measurement of the spins living at the odd sites of the string reduces the entanglement entropy of the two spins living on the edge of the string. Such an entropy reduction after measurement is accompanied by the emergence of long-range entanglement between two distant spins. This is illustrated in Fig.~\ref{fig:mimi_gs}, the inset shows the MI between the $A$ spins before the measurement which is just zero as these spins are not correlated. Note that we can compute the measurement induced MI for much larger system sizes than we could for the MI $I(A:B)$ of the previous paragraph, because we only need construct the RDM explicitly for two spins. Hence this approach is easily scalable and also accessible in experiments.

\subsection{QP properties}\label{sec:mi}

\subsubsection{Mutual information and conditional mutual information}
\begin{figure}
\includegraphics[width=0.95\linewidth]{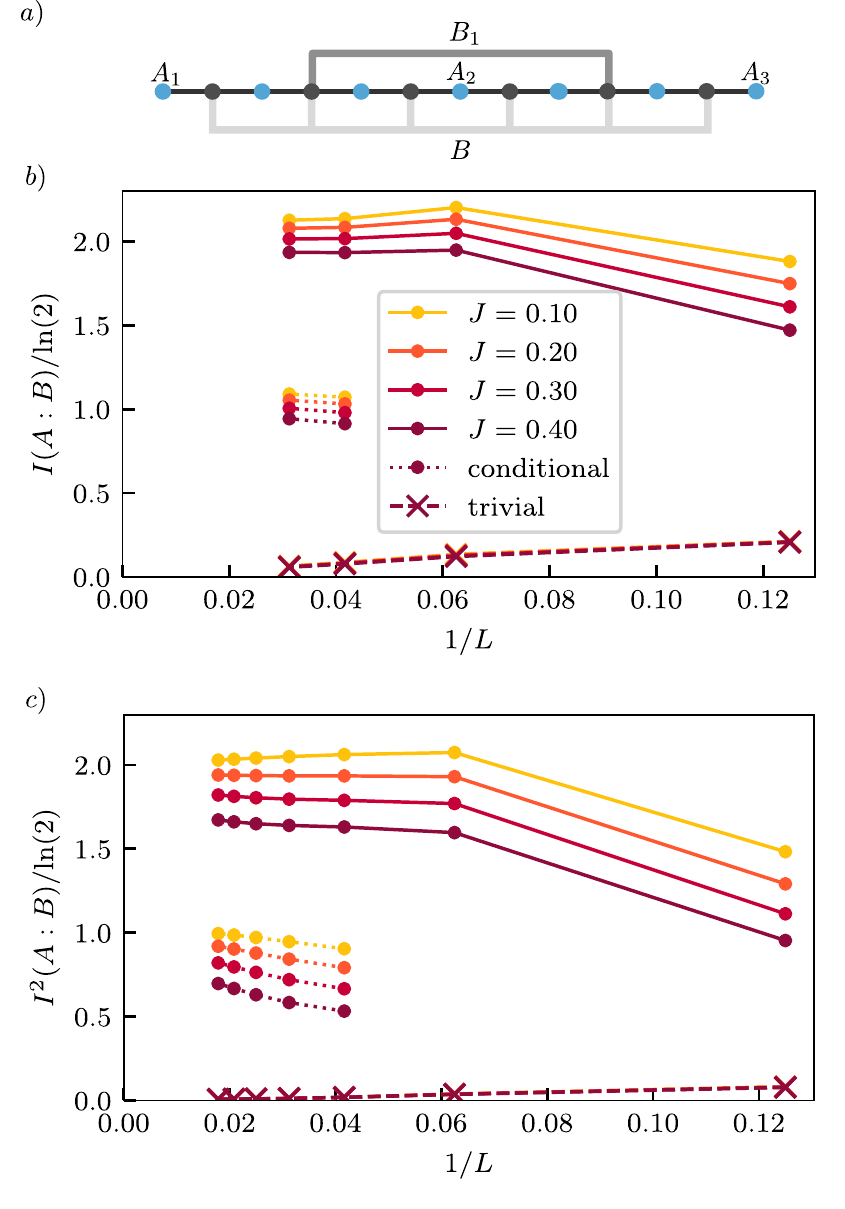}
 \caption{a) Sketch of the partitioning of the system; we take $A_1=\frac{L}{4}$, $A_2=\frac{L}{2}$ and $A_3=\frac{3L}{4}$ and system sizes divisible by eight. b) The MI in the first excited state of the cluster model. The full lines corresponds to data in the topological phase at $K'=1$,$K=0.9$ and $h=0$ for some $J$s shown in the legend, and the crosses corresponds to data in the trivial phase at $h=2.5$. The dots connected by the dotted lines are data points for the conditional MI $I(A:B_1|B_2)$, where the region $B_1$ only contains odd sites that are not directly neighboring the $A$ region. c) The second order R\'enyi MI of the first excited state at the same points in the phase diagram.} 
\label{fig:mi_qp}
\end{figure}

Now we focus on the QP state and use a slightly different model
    \begin{align} \label{eq:cluster_m}
  H=\sum_i ~& K \sigma^z_{2i}\sigma^x_{2i+1}\sigma^z_{2i+2}+K' \sigma^z_{2i-1}\sigma^x_{2i}\sigma^z_{2i+1}\nonumber\\
  &+J \sigma^z_{2i-1}\sigma^z_{2i+1} + h \sigma^x_i,
  \end{align}
 such that when $J\ll K<K'$ and $h=0$, the (non-degenerate) low-energy excitation is the energy flipping of the stabilizer term $K \sigma^z_{2i}\sigma^x_{2i+1}\sigma^z_{2i+2}$ which removes the $Z_2^a$ charge on site $2i+1$ from the domain wall between sites $2i,2i+2$~\footnote{Or it adds a $Z_2^a$ charge on site $2i+1$ provided there is no domain wall between sites $2i,2i+2$}. This can be accomplished by the QP operator $\sigma^z_{2i+1}$ which flips the charge at site $(2i+1)$. Due to the $h\sigma^z_{2i-1}\sigma^z_{2i+1}$ term, the QP acquires dynamics and can hop among odd sites as $\sum_i e^{ikR_{2i+1}}\sigma^z_{2i+1}$. Such QP unbounds the $Z_2^a$ charge with the domain wall for odd-site clusters. Consequently, the string order could vanish as the QP being inside/outside the string exactly reverse the value of string order. Nevertheless, the QP still carries non-vanishing MI between the three distant sites (living on the even lattice) labelled as $A=A_1\cup A_2 \cup A_3$ and the odd sites between them labelled as $B=B_L\cup B_R$ as in Fig.~\ref{fig:mi_qp}$a$.

  If we do not make any measurement and barely look at the QP state, region $B$ has entropy $S(B)=m\ln(2)$ with $m$ being the number of odd site spins in $B$ region. However, once we make a measurement of the spins in the $A$ region, say they are in the $(0,0,0)$ state, then  $B_L$ and $B_R$ could have all-even charge, or an even-odd/odd-even charge pattern depending on the position of the QP.  However, it is impossible to get a configuration of $B$ with odd charges on both sides based on the measurement outcome. The reason is obvious: if the QP is outside the $B$ region, then both $B_L$ and $B_R$ have even charges, but if the QP is inside the $B=B_L \cup B_R$ region, one of them has even charge and the other one odd charge. This result implies that the entanglement entropy of region $B$ can be reduced after measuring region $A$, hence their mutual information is nonzero
    \begin{align} 
I(A:B)\neq 0.
  \end{align}
This is shown in Fig.~\ref{fig:mi_qp}$b$ for some points in the phase diagram obtained by DMRG. In the simulations we take $A_1=\frac{L}{4}, A_2=\frac{L}{2}$ and $A_3=\frac{3L}{4}$ and use system sizes $L$ divisible by eight. In addition, to exclude the contribution to the MI by local correlation, we can choose a small region near $A$ and again define the conditional MI,
    \begin{align} 
I(A:B_1 | B_2)=I(A:B_1\cup B_2)-I(A:B_2).
  \end{align}
  We show some data points for the conditional MI in Fig.~\ref{fig:mi_qp}$b$, where we choose the region $B_2$ to be the four odd spins directly neighboring the $A$ spins. If we use the von Neumann entropies to compute the MIs, we are limited to small subsystems. In order to obtain data for somewhat larger systems sizes, we also computed the MIs by using the second order R\'enyi entropies as shown in Fig.~\ref{fig:mi_qp}$c$.

\subsubsection{Measurement induced non-local entanglement}

\begin{figure}
\includegraphics[width=0.95\linewidth]{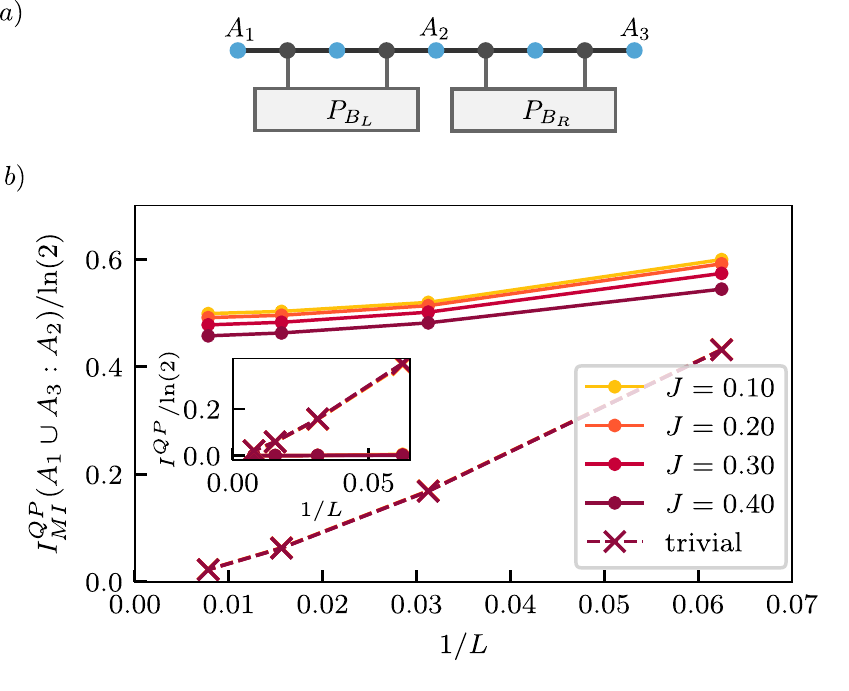}
\caption{ a) Sketch of the projection of the odd sites in between $A_1$,$A_2$ and $A_3$ to a sector with definite (e.g. even) parity.  b) The measurement-induced MI in the first excited state of the cluster model $P_{B_{L}}P_{B_R}|QP\rangle$. The full line corresponds to data in the topological phase at $K'=1$,$K=0.9$ and $h=0$ for some $J$ shown in the legend, while the dotted lines corresponds to data in the trivial phase at $h=2.5$. The inset shows the MI without the projection of the regions $B_L$ and $B_R$ to the even sector. Clearly the MI vanishes in both cases in the thermodynamic limit. } 
\label{fig:mimi_qp}
\end{figure}

Also for the QP state, we can extend our understanding of the MI by measurement induced long-range entanglement between distant spins. Of course, if we obtain the RDM of the three distant sites $A_1,A_2,A_3$ in the QP state, the RDM can be approximated as the product of the RDM of the three spins provided their distances are much larger than the correlation length. This also implies that there is no non-local entanglement among these three spins and that their mutual information vanishes, see the inset in Fig.~\ref{fig:mimi_qp}.

However, if we make a measurement of the QP state by projecting the spins in the $B$ region such that both $B_L$ and $B_R$ carry an even number of charges, the projected QP state $P_{B_L}P_{B_R}|QP\rangle$ contains non-vanishing MI (see Fig.~\ref{fig:mimi_qp}) as,
    \begin{align} 
&I(A_1\cup  A_3:A_2) \nonumber\\
&=S(A_1\cup A_3)+S(A_2)-S(A_1\cup A_3\cup A_2)\nonumber\\
&=3 \ln(2)-S(A_1\cup A_3\cup A_2).
  \end{align}
Following a similar argument from the previous paragraph, it is not hard to conclude that provided both $B_L$ and $B_R$ carry even number of charges, the possible patterns for $(A_1,A_2,A_3)$ are $(0,0,0),(1,1,1)$ for QP outside and $(1,0,0),(0,0,1),(0,1,1),(1,1,0)$ for QP inside so the total entropy of $(A_1,A_2,A_3)$ is smaller than the sum of each individual entropy $S(A_1\cup A_3)+S(A_2)=3 \ln(2)$.
This indicates that the three distant spins in $P_{B_L}P_{B_R}|QP\rangle$ contain non-vanishing long-range MI $I(A_1\cup A_3:A_2)$ as shown in Fig.~\ref{fig:mimi_qp}. Such `emergent long-range mutual information' and non-local entanglement is absent for the QP wave function due to its short-range correlations. However, the charge projection of the $B$ region entangles the three distant spins and creates non-vanishing long-range MI $I(A_1\cup A_3:A_2)$ due to the hidden correlation between the total $Z^a_2$ charges living on the odd sites of the $B$ region and the spin coherence patterns of $A_1, A_2, A_3$. \\

\begin{figure}
\includegraphics[width=0.95\linewidth]{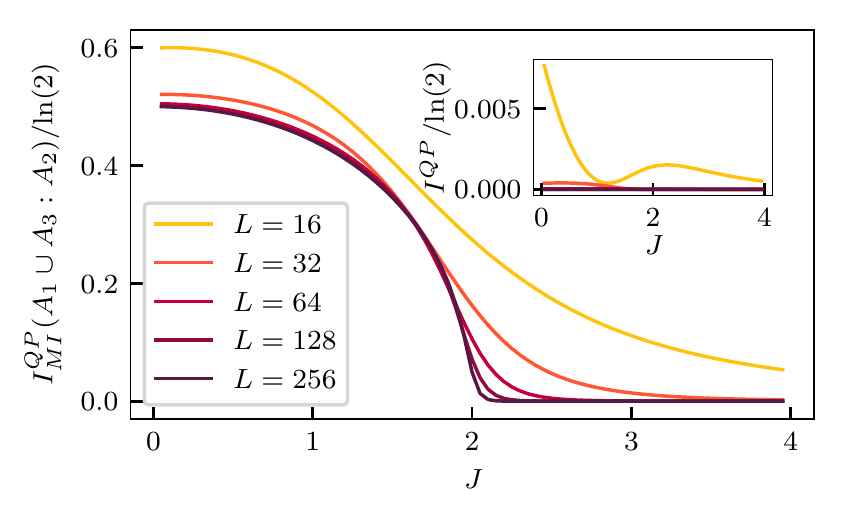}
\caption{The measurement induced MI in the first excited state of the cluster model $P_{B_L}P_{B_R}|QP\rangle$ as a function of $J$ with $K=K'=1$ and $h=0$. The inset shows the data without the projection of the regions $B_L$ and $B_R$ to the even sector.} 
\label{fig:mimi_qp_J}
\end{figure}

To explicitly illustrate that the measurement induced MI can also probe phase transitions in the QP state, we consider the case where we vary the $J$ parameter in the cluster model~\eqref{eq:cluster_m} and fix $K=K'=1$ and $h=0$. This drives the system from an SPT phase ($J<2$) to a symmetry broken phase ($J>2$), we illustrate this in Fig.~\ref{fig:mimi_qp_J} via the QP measurement induced MI. This measurement induced long-range MI is easy to implement in cold atom experimental setups~\cite{PhysRevLett.109.020504,Kaufman794,PhysRevLett.109.020505,PhysRevLett.120.050406,choo2018measurement,Brydges2019} and hence provides insight into visualizing the low energy excitation and suggests an experiment-feasible protocol to probe QPs in SPT phases.

\section{Conclusions and outlook}

In this work, we uncovered entanglement spectrum features of different QP states in various phases, and demonstrated how to identify distinct phases of matter via their low energy excitations. We anticipate the results obtained in this paper will help the exploration of quantum or thermal phases from a quantum information perspective. In particular, we expect that the entanglement properties of quasiparticle states can be adapted to examine quantum critical phenomena as well as intrinsic topologically ordered phases with fractionalized excitations, and provide a general framework for the entanglement spectroscopy of quantum and thermal phases.

We also expect that the exploration of QP entanglement and conditional MI can be extended to QP in higher dimensional systems with intrinsic topological excitations. We will report our study of the entanglement of fractionalized quasiparticles in a forthcoming manuscript. Additionally, a popular line of research relates many properties of non-equilibrium states and thermalization processes to the evolution of the entanglement structure or entanglement growth~\cite{bardarson2012unbounded}. The investigation of non-equilibrium properties of quasiparticle entanglement will be included in our forthcoming work~\footnote{Y You et al., Work in progress}.

{\em Acknowledgments.---}We would like to thank Ruben Verresen for stimulating discussions. Our DMRG simulations were performed using the TeNPy Library~\cite{Hauschild2018}. FP is funded by the European Research Council (ERC) under the European Unions Horizon 2020 research and innovation program (grant agreement No. 771537). FP acknowledges the support of the Forschungsgemeinschaft (DFG, German Research Foundation) under Germany's Excellence Strategy EXC-2111-390814868. FP and EW  acknowledge the DFG TRR80. YY and FP acknowledge the support from the Banff International Research Station, where this work was partly initiated. SLS acknowledges support from the United States Department  of  Energy  via  grant  No.   DE-SC0016244. Additional support was provided by the Gordon and Betty Moore Foundation  through  Grant  GBMF8685  towards  the Princeton theory program.
 
\bibliography{biblio.bib}
\end{document}